\documentclass[twocolumn]{aastex63}
\usepackage{mathtools}
\usepackage{amsmath}

\submitjournal{ApJL}

\shorttitle{Magnetic field amplification in shock precursors}
\shortauthors{Peterson, Glenzer, \& Fiuza}

\begin{document}

\title{Magnetic field amplification by a plasma cavitation instability in relativistic shock precursors}

\author[0000-0001-7730-5724]{J. R. Peterson}
\email{jrpete@stanford.edu}
\affiliation{SLAC National Accelerator Laboratory, 2575 Sand Hill Rd, Menlo Park, CA 94025}
\affiliation{Physics Department, Stanford University, 450 Serra Mall, Stanford, CA 94305}
\author{S. Glenzer}
\author[0000-0002-8502-5535]{F. Fiuza}
\email{fiuza@slac.stanford.edu}
\affiliation{SLAC National Accelerator Laboratory, 2575 Sand Hill Rd, Menlo Park, CA 94025}

\begin{abstract}

Plasma streaming instabilities play an important role in magnetic field amplification and particle acceleration in relativistic shocks and their environments. However, in the far shock precursor region where accelerated particles constitute a highly relativistic and dilute beam, streaming instabilities typically become inefficient and operate at very small scales when compared to the gyroradii of the beam particles. We report on a plasma cavitation instability that is driven by dilute relativistic beams and can increase both the magnetic field strength and coherence scale by orders of magnitude to reach near-equipartition values with the beam energy density. This instability grows after the development of the Weibel instability and is associated with the asymmetric response of background leptons and ions to the beam current. The resulting net inductive electric field drives a strong energy asymmetry between positively and negatively charged beam species. Large-scale particle-in-cell simulations are used to verify analytical predictions for the growth and saturation level of the instability and indicate that it is robust over a wide range of conditions, including those associated with pair-loaded plasmas. These results can have important implications for the magnetization and structure of shocks in gamma-ray bursts, and more generally for magnetic field amplification and asymmetric scattering of relativistic charged particles in plasma astrophysical environments.

\end{abstract}

\section{Introduction} \label{sec:intro}

Relativistic streaming plasma instabilities are important in a wide variety of energetic astrophysical environments such as gamma-ray bursts (GRBs), supernova remnants shocks, and blazar jets. These objects can produce relativistic charged particles through different processes including nonthermal particle acceleration in collisionless shocks, photon-photon collisions, and electron-positron (pair) cascades. In weakly magnetized plasmas, as typically associated with relativistic shocks in GRBs and other jet environments, these particle beams drive plasma instabilities that play a very important role in the amplification of magnetic fields, strongly influencing particle scattering, acceleration, and radiation emission.

Plasma microinstabilities, such as the Weibel (or current filamentation) instability \citep{weibel59,fried59}, have attracted significant attention as leading mechanisms for the rapid amplification of magnetic fields \citep{medvedev99}. Kinetic particle-in-cell (PIC) simulations have shown that the Weibel instability is important in the formation of relativistic collisionless shocks and nonthermal particle acceleration \citep{silva03,spitkovsky07,spitkovsky08,martins09,sironi13,lemoine19PRL}. However, plasma microinstabilities typically saturate at small, plasma skin depth scales. These kinetic scales are much smaller than the magnetic coherence length required to explain polarized GRB emission \citep{gruzinov99,covino99,steele09,gill20}, and the rapid decay of such small-scale fields in the downstream is at odds with inferred downstream GRB field strengths \citep{chang08,keshet09,lemoine15}.

Previous numerical studies have primarily considered the case of symmetric streaming plasmas or beams. However, in most scenarios of interest, such as in the precursors of relativistic shocks and in blazar jets the beam-plasma systems are highly asymmetric, with a relativistic, hot, and dilute beam propagating on a cold and dense background plasma. The few existing studies in this regime \citep{sironi14} show that microinstabilities become very inefficient and saturate at very low magnetization levels $\epsilon_B < 10^{-3}$ (where $\epsilon_B$ is the ratio of the magnetic energy density to the beam kinetic energy density). Furthermore, $\gamma\gamma$ collisions (and associated pair cascades) can load the shock precursor in GRB environments with electron-positron pairs \citep{thompson00,meszaros01,beloborodov02,ramirez-ruiz07} and it is not clear how pair loading will impact the long-term nonlinear evolution of the instabilities, the resulting magnetization, and particle acceleration in the shock.

In this Letter, we show that dilute relativistic beams propagating on an electron-ion (or pair-ion) background can give rise to a nonlinear plasma instability that exponentially amplifies both the strength and coherence length of the magnetic field. The instability arises after the saturation of the Weibel instability and is driven exclusively by the beam electrons, regardless of the beam positron/ion composition, as they are charge- but not current-neutralized by background ions. The asymmetric response of the background species leads to an energy asymmetry between the beam species, as only the beam electrons are inductively decelerated by the increasing magnetic field strength. We present analytical predictions for the growth and saturation level of the instability, which are validated by two-(2D) and three-dimensional (3D) PIC simulations.

\section{Setup}
We explore the nonlinear late-time evolution of beam-plasma systems using 2D and 3D fully kinetic simulations with the relativistic PIC code OSIRIS \citep{fonseca02,fonseca08}. We consider the general case of a dilute, relativistic pair-ion beam propagating in a cold pair-ion background plasma with initial beam-to-background plasma density ratio $\alpha = n_{b}/n_{0} \ll 1$ (the indices $b$ and $0$ denote beam and background quantities). The beam species are initially in equipartition, with beam leptons having Lorentz factor $\gamma_{be} \gg 1$ (corresponding to an initial velocity $v_{be} \sim c$) and beam ions having $\gamma_{bi} = [1-(v_{bi}/c)^2]^{-1/2} = 1+m_e(\gamma_{be}-1)/m_i$, with $m_e$ and $m_i$ the lepton and ion mass, respectively. The system is initialized as charge and current neutral with $n_{0e^-} = (1+Z_\pm) n_{0i}$, $n_{0e^+} = Z_\pm n_{0i}$, $n_{be^-} = \alpha (1+Z_\pm) n_{0i}$, $n_{be^+} =  \alpha Z_\pm n_{0i}$, $n_{bi} =  \alpha n_{0i}$, $v_{0e^-} = - v_{0e^+} = \alpha (c-v_{bi})/(1+2 Z_\pm)$, and $v_{0i} = 0$, where $Z_\pm$ is the pair loading factor and indices $e^-$, $e^+$, and $i$ refer to electrons, positrons, and ions.

The beam propagates in the $x$-direction and the typical domain size of the simulations is $4000\times4000~(c/\omega_p)^2$ in 2D $yz$, $4000\times3000~(c/\omega_p)^2$ in 2D $xy$, and $6000\times1400\times1400~(c/\omega_p)^3$ in 3D. The cell size 
was varied between $\Delta = 0.0625-1.0~c/\omega_p$, where $c$ is the speed of light, $\omega_p = [4 \pi (1+Z_\pm) n_{0i} e^2 / m_e]^{1/2}$ is the background electron plasma frequency, and $-e$ the electron charge. All simulations use a realistic mass ratio $m_i/m_e = 1836$. The time step is chosen according to the Courant–Friedrichs–Lewy condition, and we use 4 particles per cell in 2D (8 in 3D) per species. We have tested different simulation box sizes, resolutions, and numbers of particles per cell to ensure convergence of the results and have used a third order particle interpolation scheme for improved numerical accuracy.

\section{Results}
\subsection{Transverse dynamics}

\begin{figure}[]
\centering
\includegraphics[width=\linewidth]{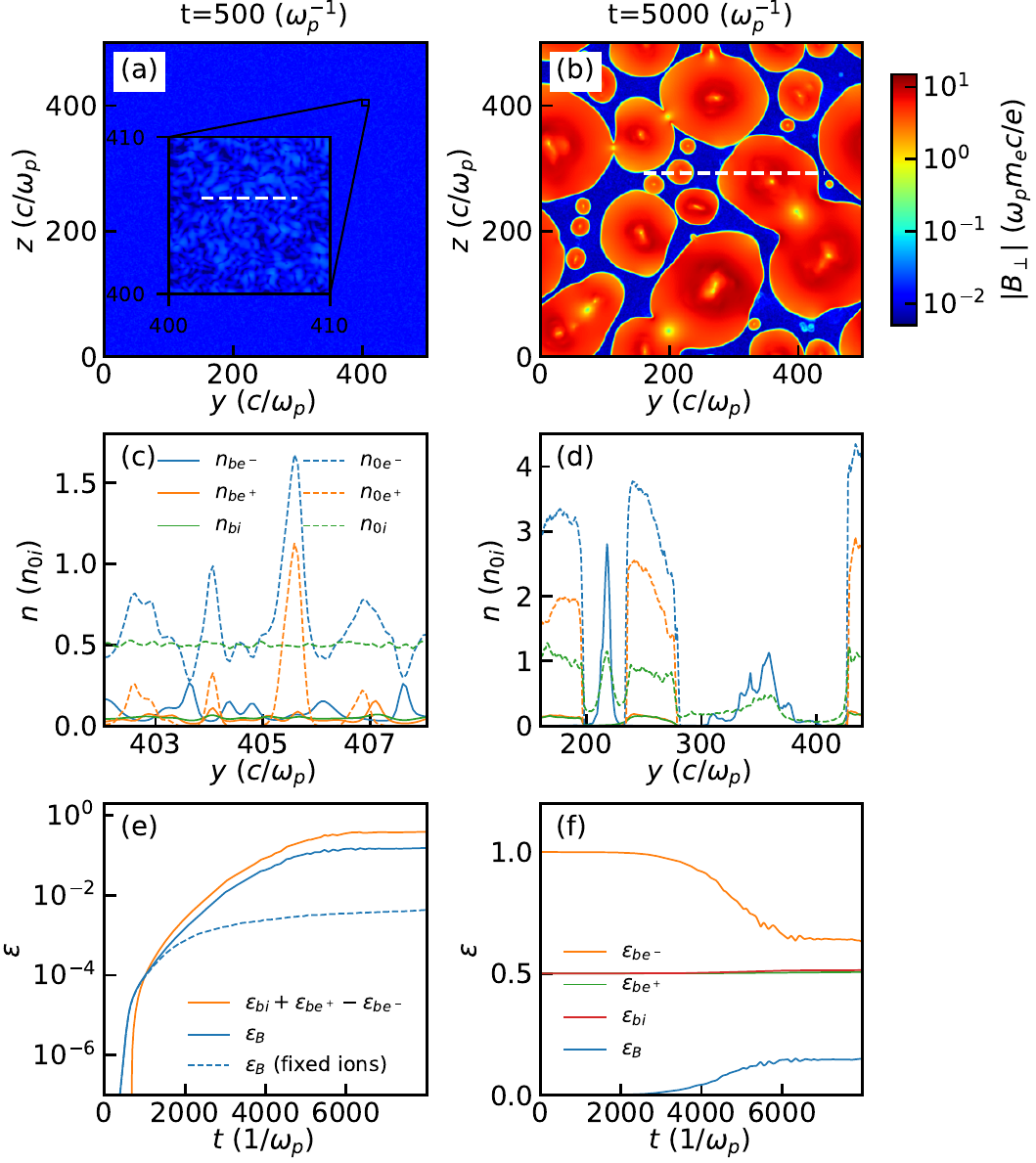}
\caption{\label{fig:2d_transverse}
2D simulation of the transverse dynamics of a dilute, relativistic pair-ion beam with $\alpha=0.1$ and $\gamma_{be}=1000$ propagating into a cold pair-ion plasma. The pair loading factor is $Z_\pm = 1$ and the beam propagates into the page. Magnetic field profiles (a,b) and density lineouts (c,d) are shown at (a,c) $t=500~\omega_p^{-1}$, the time of saturation of the Weibel instability, and (b,d) $t=5\times10^5~\omega_p^{-1}$, the time of saturation of the cavitation instability. The lineouts are taken along the white dashed lines in (a,b). The energy evolution of the system is reported in (e) log scale and (f) linear scale.
}
\end{figure}

We start by considering the transverse dynamics of the beam-plasma interaction. In Fig. \hyperref[fig:2d_transverse]{\ref*{fig:2d_transverse}} we show results from a 2D simulation in the $yz$ plane (perpendicular to the beam propagation) with a cold pair-ion beam ($\alpha = 0.1$,  $\gamma_{be} = 1000$, $Z_\pm=1$) in a cold background plasma, which is representative of the dominant dynamics observed in our simulations. We observe a first, rapid phase of magnetic field amplification that terminates at $t \sim 500~\omega_p^{-1}$ and corresponds to the well-established Weibel instability. The measured growth rate $\Gamma=1.31\times10^{-2}\omega_p$ is in good agreement with the theoretical Weibel growth rate $\Gamma_W = \sqrt{2\alpha/\gamma_{be}}\omega_p = 1.41\times10^{-2}\omega_p$. The resulting magnetic field has a very small spatial scale of the order of the electron skin depth $c/\omega_p$ of the background plasma [Figs. \hyperref[fig:2d_transverse]{\ref*{fig:2d_transverse}(a)} and \hyperref[fig:2d_transverse]{\ref*{fig:2d_transverse}(c)}] as expected from linear theory \citep{silva02}. The saturation level corresponds to a magnetization of $\epsilon_B = B^2/(8\pi n_{b0} \gamma_{be} m_e c^2) \sim 4\times10^{-5}$, which is also close to the theoretical value expected due to magnetic trapping, $\epsilon_B \sim \alpha/(2\gamma_{be}) \sim 5 \times 10^{-5}$ \citep{davidson72}. At this stage the background ions did not yet have time to respond and if they are artificially kept fixed (immobile, neutralizing species), we observed that after saturation the magnetic energy only slightly increases due to filament merging and compression \citep{honda00pop} (Fig. \hyperref[fig:2d_transverse]{\ref*{fig:2d_transverse}(e)}, dashed line); the final magnetization remains at the $\epsilon_B \sim 10^{-3}$ level.

The dynamics change significantly on longer time scales associated with the background ion motion. A second stage of magnetic field growth is clearly visible in Fig. \hyperref[fig:2d_transverse]{\ref*{fig:2d_transverse}(b,d,e)}, which saturates at approximately $t=5\times 10^3~\omega_p^{-1}$. During this phase, large density cavities form in the background plasma [Fig. \hyperref[fig:2d_transverse]{\ref*{fig:2d_transverse}(d)}] and both the energy and wavelength of the magnetic field are amplified by another two orders of magnitude to $\epsilon_B=0.15$ and $\lambda_B \sim 200 c/\omega_p \sim r_{gb}$ [Fig. \hyperref[fig:2d_transverse]{\ref*{fig:2d_transverse}(b)}], respectively, where $r_{gb}\sim\gamma_{be}m_e c^2/(eB_\perp)$ is the beam electron gyroradius.

Simultaneously with the magnetic field amplification, an important energy asymmetry develops; the evolution of the beam electron kinetic energy, normalized to its initial value, $\epsilon_{be^-}=\gamma_{be^-}/\gamma_{be}$ in Fig. \hyperref[fig:2d_transverse]{\ref*{fig:2d_transverse}(f)} drops by nearly a third, while the normalized beam positron and ion energies,  $\epsilon_{be^+}$ and $\epsilon_{bi}$, are almost unchanged.

We find that this second growth phase corresponds to a nonlinear electron streaming instability recently discovered in dilute relativistic electron beams \citep{peterson21} that arises after the saturation of the Weibel instability. In the present case, the magnetic pressure around the current filaments expels the background leptons, resulting in a space charge field that will pull most of the background ions out to restore quasi-neutrality, forming a small cavity in the background plasma. Beam positron/ion filaments are charge- and current-neutralized by a small residual density $\sim n_{be^-}$ of background electrons. However, beam electron filaments are charge-neutralized by background ions, which cannot effectively screen their current due to large inertia [Fig. \hyperref[fig:2d_transverse]{\ref*{fig:2d_transverse}(d)}]. The magnetic pressure in the unscreened beam electron cavities causes them to expand, exposing more current and leading to exponential growth in both the cavity size and magnetic field strength [Fig. \hyperref[fig:2d_transverse]{\ref*{fig:2d_transverse}(b)}].

The formation of small plasma density cavities has been observed in previous simulations of collisionless shocks \citep{fiuza12,ruyer15,naseri18}. However, previous simulations have typically used either much reduced ion to electron mass ratios or small domains in upstream region, and as such have not recognized the growth of an instability related to the cavities. Below we describe and characterize in detail the development, growth, and saturation of this cavitation instability for the general case of pair-ion beam-plasma systems.

\subsection{Growth rate and saturation of cavitation instability}

In order to calculate the growth rate of the cavitation instability we consider an ultrarelativistic, dilute pair-ion beam ($\gamma_{be} \gg 1$ and $\alpha \ll 1$) with $m_i \gg m_e$. We use as a starting point the saturation phase of the Weibel instability. At this stage, alternating beam current filaments are produced that expel most of the background plasma in the filament region, forming a cavity of diameter $\lambda_B$, which is the magnetic wavelength. In the beam electron cavities, the magnetic pressure must expel all species except the beam electrons and a charge-neutralizing population of background ions given by $n_{0i} \sim n_{be^-}$ if $\alpha Z_\pm \leq 1$ and $n_{0i}$ otherwise. Thus the total relativistic mass density which must be expelled and builds up at the cavity wall is
\begin{align}
\begin{split}\label{eq:rho}
    \rho_w = \ &n_{0i} \big\{m_i\ \textrm{max}\{1- \alpha Z_\pm,\alpha\} \\&+ m_e \left[1+2Z_\pm+\alpha\gamma_{be}(Z_\pm+1) \right]\big\},
\end{split}
\end{align}
where max$\{a,b\}$ is the greater of $a$ and $b$ and we have neglected corrections of order $\alpha$ (see Appendix \ref{appendix:growthrate} for details).

In the limit $\alpha Z_\pm \leq 1$, the background ions are able to completely charge neutralize the beam electrons. The unscreened beam electron current $J_b \approx -en_{be^-}v_{be}$ gives rise via Amp{\`e}re's law to a magnetic field $B=\alpha \beta_{be} \lambda_B \omega_p^2 m_e/(2e)$ and magnetic pressure $P_B = B^2/(8\pi)$ at the cavity walls, where $\beta_{be}=v_{be}/c$. In the opposite limit $\alpha Z_\pm > 1$, the background ions cannot completely screen the beam electron charge; the net charge density $e (n_{0i}-n_{be^-})$ in the cavity produces a radial electric field $E_r = -\alpha\lambda_B \omega_p^2 m_e(1 - \alpha Z_\pm)/(2 Z_\pm e)$. In this regime, beam positrons/ions dominate the wall inertia and this attractive electric field reduces the net force on the cavity wall by a factor $\alpha Z_\pm$. We account for this with an effective pressure valid in both regimes
\begin{equation}
    P_{\textrm{eff}} = P_B\ \textrm{min}\left\{1,1/(Z_\pm\alpha)^2\right\}
\end{equation}
where min$\{a,b\}$ is the lesser of $a$ and $b$.

In slab geometry, the wall has mass $m_w \approx \rho_w A \lambda_B/2$ for arbitrary area $A$. The wall momentum is $p_w = m_w d(\lambda_B/2)/dt$ and will increase under the effective pressure according to $dp_w/dt = P_{\textrm{eff}}A$ which can be written as
\begin{equation}
  \frac{d}{d t}\left(\lambda_B \frac{d\lambda_B}{dt}\right) = \frac{\alpha^2 \beta_{be}^2}{2}\frac{n_{be^-}\ \textrm{min}\left\{1,1/(Z_\pm\alpha)^2\right\}}{\rho_w} \lambda_B^2 \omega_p^2.
\end{equation}
The solution is exponential growth of the form $\lambda_B(t) = \lambda_{B0} e^{\Gamma t}$ with a rate
\begin{equation}\label{eq:gr1}
  \frac{\Gamma}{\omega_p} = \alpha \beta_{be} \textrm{min}\left\{1,1/(Z_\pm\alpha)\right\} \sqrt{\frac{n_{0e^-} m_e}{\delta \rho_w}}
\end{equation}
where $\delta = 4$ in the slab geometry, and a similar calculation for cylindrical geometry yields $\delta = 3$.

In the limit $Z_\pm \ll 1$, the growth rate reduces to
\begin{equation}\label{eq:gr2}
  \frac{\Gamma}{\omega_p} = \alpha \beta_{be} \sqrt{\frac{m_e}{\delta (m_i + m_e\alpha \gamma_{be})}}
\end{equation}
which is equivalent to that of a pure electron-ion or electron-positron beam on an electron-ion background. Moreover, when the background ions dominate the wall mass ($\alpha \gamma_{be}m_e \ll m_i$), the growth rate reduces to the pure electron beam case in \cite{peterson21}.

\begin{figure}[]
\centering
\includegraphics[width=\linewidth]{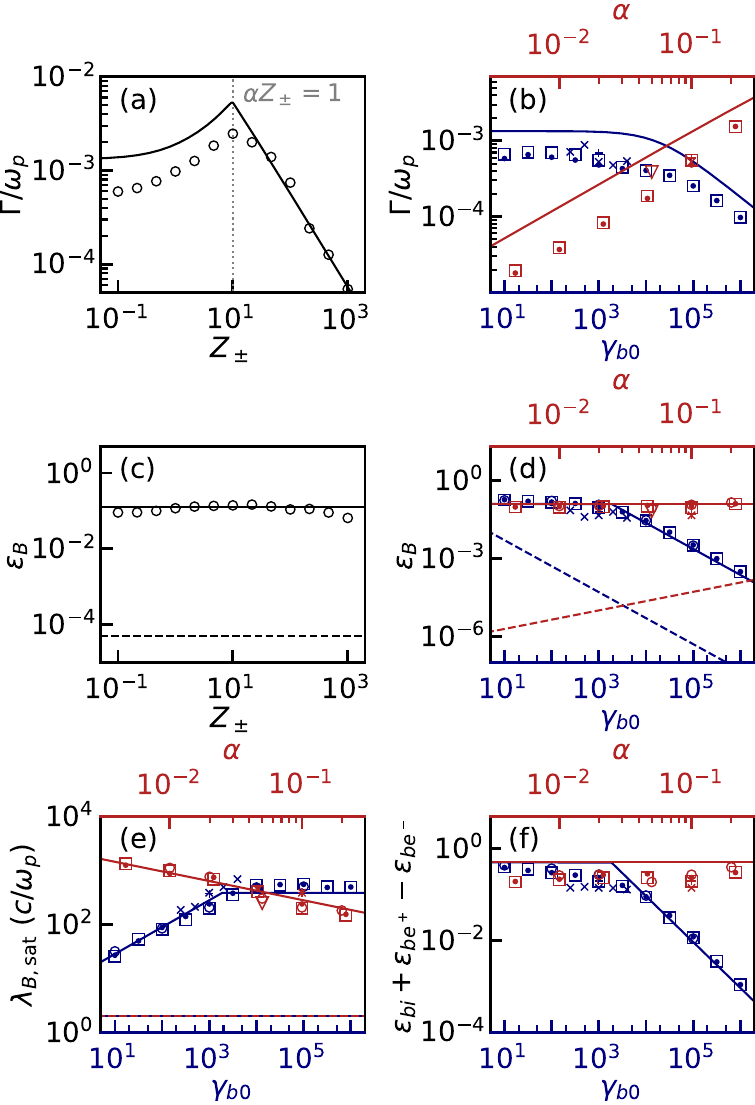}
\caption{\label{fig:scalings}
Comparison between analytical growth rates and saturation values for the cavitation instability (solid lines), Weibel instability (dashed lines), and PIC simulations. Results from 2D simulations of the transverse $yz$-plane are shown for variable pair loading factor $Z_\pm \neq 0$ (open circles), pure electron-ion ($Z_\pm=0$) beam-plasma (filled circles), and pure electron-positron beam on electron-ion plasma (squares). Simulations capturing the longitudinal dynamics are shown for an electron-positron beam propagating on electron-ion plasma in 3D (triangles) and 2D $xy$-plane geometry for unmagnetizd ($\times$) and magnetized ($+$) ($\sigma_\perp \sim 10^{-6}$) initial conditions. (a,b) Cavitation instability growth rate. (c,d) Saturation magnetic field wavelength. (e) Saturation magnetization. (f) Beam energy asymmetry. All parameter scans use as fixed parameters $\alpha = 0.1$ and $\gamma_{be} = 1000$.
}
\end{figure}

The growth rate in Eq. \ref{eq:gr1} is verified over a wide range in $\gamma_{be}$, $\alpha$, and $Z_\pm$ by 2D $yz$-plane simulations. As predicted, the growth rate in Fig. \hyperref[fig:scalings]{\ref*{fig:scalings}(a)} is maximized at $\alpha Z_\pm=1$, for which background ions can charge-neutralize the beam electrons without contributing to the cavity wall mass. The dependence on $\gamma_{be}$ and $\alpha$ is explored with simulations in Fig. \hyperref[fig:scalings]{\ref*{fig:scalings}(b)} for $Z_\pm=0,1$, which clearly show the transition between background ion-dominated cavity wall mass for $\gamma_{be} < m_i/(\alpha m_e)$ and beam positron/ion-dominated wall mass for $\gamma_{be} > m_i/(\alpha m_e)$. The scaling with both $\gamma_{be}$ and $\alpha$ is in good agreement with the theory. Pair beams propagating in electron-ion plasma are shown to behave nearly identically to electron-ion beams as expected from Eq. \ref{eq:gr1}. The reduction in the growth rate by $\sim50\%$ in some simulations is due to competition between cavities, which lowers the pressure drop across the wall.

Growth of the cavities and magnetic field amplification will saturate when either the beam electrons or the background ions can respond to reduce the net current in the cavity. The first case occurs when the beam electron gyroradius becomes comparable to the cavity radius $\lambda_B/2$. The second case occurs when the background ions in the cavity are accelerated by the inductive electric field $E_x = \alpha \Gamma (\lambda_B/2)^2 (m_e \omega_p^2/ce)$ and neutralize the beam electron current. Combining the two criteria \citep{peterson21} leads to the saturation cavity size and magnetic wavelength
\begin{equation}\label{eq:sat_length}
  \lambda_{B,\rm{sat}} \sim \sqrt{\frac{8}{\alpha} \textrm{min}\left\{\gamma_{be},\frac{m_i}{m_e}\right\}} \frac{c}{\omega_p}.
\end{equation}

The saturation magnetization is estimated as $\epsilon_B = B_{\rm{sat}}^2/(16\pi n_{be} \gamma_{be} m_e c^2)$, where a factor of 1/2 is included to take into account that cavities will occupy roughly only half of the system volume. The average magnetic field is $B_{\rm{sat}}\sim\alpha \lambda_{B,\rm{sat}} \omega_p^2 m_e/(4e)$, which results in
\begin{equation}\label{eq:sat_eps}
  \epsilon_B \sim \frac{1}{8} \textrm{min}\left\{1,\frac{m_i}{\gamma_{be} m_e}\right\}.
\end{equation}

Lastly, we estimate the energy asymmetry between the electrons and positrons/ions, which arises from the inductive electric field in the cavities. By multiplying this electric force by the distance traveled during one growth period, we estimate the work on a beam electron in the cavity $W = -e E_x c/\Gamma = -2\  \textrm{min}(\gamma_{be},m_i/m_e) m_e c^2$. However, the average beam electron will experience $\langle W \rangle \sim W/4$ since only about half of the system volume contains cavities and only about half of those beam electrons were in the cavity during an e-folding growth of the cavity expansion. This yields an electron-positron/ion energy asymmetry
\begin{equation}\label{eq:sat_asymmetry}
  \epsilon_{bi}+\epsilon_{be^+}-\epsilon_{be^-} \sim \frac{1}{2} \textrm{min}\left\{1,\frac{m_i}{\gamma_{be} m_e}\right\}.
\end{equation}

Interestingly, we see that the magnetization level, magnetic wavelength, and beam energy asymmetry do not depend on the details of the beam composition and pair loading factor in the regime considered here.
The predictions in Eqs. \ref{eq:sat_length} -- \ref{eq:sat_asymmetry} are verified by 2D $yz$-plane simulations over a wide range of parameters, as illustrated in Figs. \hyperref[fig:scalings]{\ref*{fig:scalings}(c)--(f)}. Our results demonstrate the transition between the two different saturation mechanisms at $\gamma_{be} \sim m_i/m_e$. The values of $\lambda_{B,\rm{sat}}$ and $\epsilon_B$ reached by the cavitation instability are orders of magnitude larger than those reached by the Weibel instability in this dilute beam regime. Moreover, we confirm that the cavitation instability leads to a large energy asymmetry between beam species.

\subsection{Longitudinal dynamics and beam temperature}

So far, our analysis considered only the transverse evolution of the system under relatively idealized conditions of a cold, uniform beam. Astrophysical beams are typically relativistically hot. Electron-positron beams from $\gamma\gamma$ collisions have comoving temperatures of order $T_b^\prime \sim m_e c^2$ since the  pair-production cross section peaks  slightly above the threshold   energy. Similarly, in simulations of particle acceleration in collisionless shocks, comoving lepton beam temperatures $T_b^\prime \gtrsim m_e c^2$ are commonly inferred \citep{spitkovsky08,lemoine19PRL}. We thus consider here a beam with drifting Maxwell-J{\"uttner} lepton distributions with comoving temperatures of order $T_b^\prime \sim m_e c^2$. When boosted to the laboratory frame, this yields longitudinal and transverse momentum spreads of $\Delta P_{||} \sim \gamma_{be} m_e c$ and $\Delta P_\perp \sim m_e c$, respectively, which will affect differently the growth of longitudinal and transverse modes.

In the cold limit, electrostatic modes, such as the oblique instability \citep{bret10}, are typically the fastest growing modes in the dilute, relativistic regime and can heat the beam and background plasma before saturating. This may affect the growth of the cavitation instability as it requires that the magnetic pressure $P_B$ (due to the Weibel instability) exceeds the thermal pressure $P_{th}$ (due to the electrostatic modes) locally at the current filaments. The large $\Delta P_{||}$ of relativistic beams will impact the growth of the electrostatic modes. In particular, it will stabilize their `quasilinear relaxation' phase and greatly reduce $P_{th}$ when compared to the cold limit, as discussed in \cite{sironi14}.

\begin{figure}[]
\centering
\includegraphics[width=\linewidth]{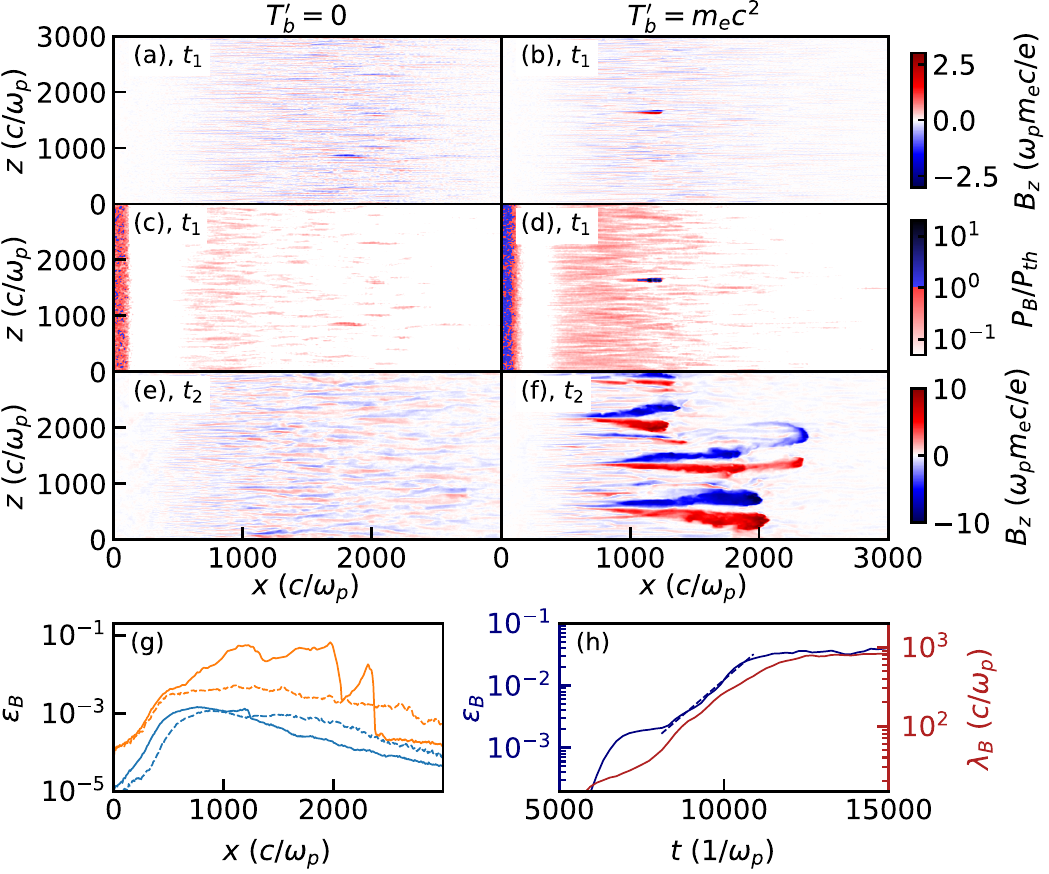}
\caption{\label{fig:semi_infinite}
Comparison of 2D simulations of the long-term magnetic field amplification for a cold (a,c,e) and a hot beam with comoving temperature $T_b=m_e c^2$ (b,d,f,h). The dilute ($\alpha=0.1$) electron-positron (pair) beam has mean Lorentz factor $\langle \gamma_{be} \rangle = 1000$ in both cases and propagates in the positive $x$-direction through an electron-ion plasma. Magnetic field profiles (a,b) and pressure ratio $P_B/P_{th}$ (c,d) are taken at $t_1=3000~\omega_p^{-1}$ when the Weibel instability saturates. Magnetic field profiles in (e,f) are taken at $t_2=12000~\omega_p^{-1}$ when the cavitation instability saturates. The $z$-averaged magnetization is shown in (g) at $t_1$ (blue) and $t_2$ (orange) for the cold (dashed) and hot (solid) beam. The evolution of the magnetization and magnetic coherence length are reported in (h) for the hot beam.
}
\end{figure}

For $\Delta P_\perp/(\gamma_{be} m_e c) \sim 1/\gamma_{be} \ll 2\alpha$, the growth and saturation level of the Weibel instability is not significantly affected by the beam temperature \citep{silva02} and we recover the cold limit discussed in the previous section. Thus we expect that relativistic beam temperatures will be important and overall aid the growth of the cavitation instability.

We have performed 2D simulations in the $xy$-plane to study how the beam temperature and longitudinal modes affect the growth of the cavitation instability. We illustrate here the case of a pure pair beam propagating on an electron-ion background plasma for simplicity. In Fig. \ref{fig:semi_infinite} we compare both a cold ($T_b^\prime = 0$) and hot ($T_b^\prime = m_e c^2$) pair beam which enters from the left side of the simulation box at $t=0$. The use of a finite beam and open boundary conditions in $x$ (periodic in $y$) is important to avoid numerical artifacts inherent to fully-periodic simulations that can stabilize the growth of the instability as explained in Appendix \ref{appendix:longitudinal_simulations}. The cold background electron-ion plasma extends from $x=200~c/\omega_p$ to $x=3800~c/\omega_p$ in order to avoid unphysical fields near the boundaries.

The early-time magnetic field profile in Fig. \hyperref[fig:semi_infinite]{\ref*{fig:semi_infinite}(a,b)} shows that the Weibel instability reaches similar field strengths in each case. However, the ratio $P_B/P_{th}$ in Fig. \hyperref[fig:semi_infinite]{\ref*{fig:semi_infinite}(c,d)} is much larger on average in the hot beam case due to the stabilization of the electrostatic quasi-linear relaxation. Indeed, only in the hot beam case do we observe $P_B/P_{th}>1$ and the growth of the cavitation instability, as shown in Fig. \hyperref[fig:semi_infinite]{\ref*{fig:semi_infinite}(e,f,g)} The magnetization and saturation magnetic wavelength produced by the cavitation instability are orders of magnitude larger than the cold case. We measure the growth of these quantities in the frame of the cavitation instability, where the drift speed $v_d = \mathbf{E} \times \mathbf{B}/B^2$ vanishes, and show simultaneous exponential growth as predicted by our model [Fig. \hyperref[fig:semi_infinite]{\ref*{fig:semi_infinite}(h)}]. The growth rate, magnetic wavelength, and saturation magnetization for a series of 2D $xy$-plane simulations are plotted in Fig. \ref{fig:scalings} where they broadly match the fully-transverse geometry. Similar results were also obtained for electron-ion and pair-ion beams with varying pair loading factors.

These results indicate that the cavitation instability can play a very important role in magnetic field amplification in the precursor of relativistic shocks, where shock accelerated species are expected to be in near equipartition with each other, constituting a hot and dilute relativistic beam that propagates in the ambient plasma medium \citep{sironi13}. We note that recent work by \cite{bresci21} using fully-periodic simulations has observed a similar growth of cavities for an electron-ion beam that is not in equipartition — the ion inertia was dominant — as expected in non-relativistic or mildly relativistic shocks, but saw no growth when the beam was in equipartition. As we demonstrate in our work an asymmetry in the inertia of the beam species is not a requirement for the development of the cavitation instability — the difference in inertia naturally present between the leptons and ions in the ambient medium upstream of relativistic shocks is sufficient. Thus, we expect this instability to operate efficiently in different shock scenarios covering a wide range of Lorentz factors and beam-plasma compositions as shown in Fig. \ref{fig:scalings}.

\subsection{External magnetic field}
An external magnetic field can help stabilize the Weibel instability, and thus impact the growth of the cavitation instability, when the beam transverse deflection during one growth period, $(c/\Gamma)^2/(2r_{gb})$, exceeds the dominant wavelength, $ \lambda_{B,\rm{sat}}$. This can be written in terms of the ambient upstream magnetization as $\sigma_\perp > \alpha^2/[(\sin\theta)^2m_i/m_e]$, with $\sigma_\perp = B_{0}^2/(4\pi n_{0i} m_i c^2) \approx 5\times 10^{-11} [B_0(\mu\rm{G})]^2 [n_{0i}(\rm{cm}^{-3})]^{-1}$ and
$\theta$ the angle between the field and the beam propagation direction. For $\sigma_\perp \sim 10^{-10}$, as may be expected in some GRB conditions, this is easily satisfied for $\alpha \gtrsim 5\times10^{-4}$. We confirm this by adding a 2D $xy$-plane simulation with $\alpha=0.1$, $\gamma_{be}=1000$, $\theta=80^\circ$, and $\sigma_\perp=10^{-6}$ to Fig. \ref{fig:scalings}. The growth and saturation of the cavitation instability is very similar to the unmagnetized case.

\subsection{Three-dimensional simulations}

\begin{figure}[]
\centering
\includegraphics[width=\linewidth]{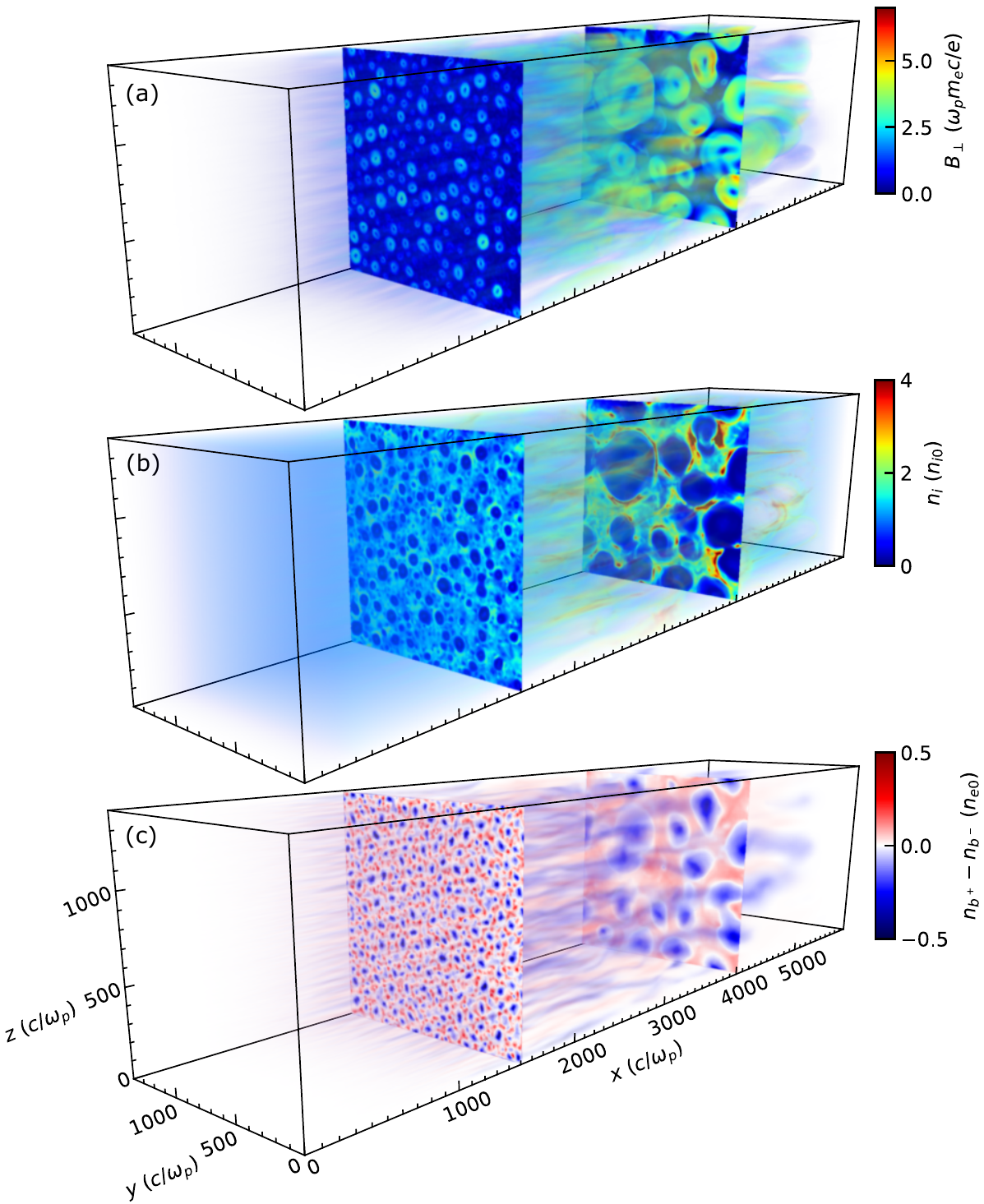}
\caption{\label{fig:3d}
Development of the cavitation instability in a 3D simulation of a dilute ($\alpha=0.05$) electron-positron (pair) beam propagating in the positive $x$-direction through a cold electron-ion plasma. The beam has comoving temperature $T_b=m_e c^2$ and mean Lorentz factor $\langle \gamma_{be} \rangle = 1000$. The transverse magnetic field amplitude (a), background ion density (b), and beam charge density (c) are reported at $t=12000~\omega_p^{-1}$, the saturation time of the cavitation instability. The opacity scales linearly with the value.
}
\end{figure}

Finally, we have confirmed that in full 3D geometry the cavitation instability dynamics is still well described by our model. In Fig. \hyperref[fig:3d]{\ref*{fig:3d}} we present the results of a 3D simulation of a hot semi-infinite pair beam entering a cold electron-ion background which extends from $x=200~c/\omega_p$ to $x=5800~c/\omega_p$. Indeed, we observe strong amplification of the magnetic field to large scales [Fig. \hyperref[fig:3d]{\ref*{fig:3d}(a)}] associated with plasma cavities [Fig. \hyperref[fig:3d]{\ref*{fig:3d}(b)}], characteristic of the cavitation instability. The dominant current in the cavities is driven by the beam electrons [Fig. \hyperref[fig:3d]{\ref*{fig:3d}(c)}]. The growth rate and saturation values are all shown to be in good agreement with the analytical predictions as seen in Fig. \ref{fig:scalings}.

\section{Conclusion}
We have shown that dilute, relativistic charged particles beams propagating on a cold and denser ambient plasma are subject to a secondary cavitation instability that operates after saturation of the Weibel instability. This cavitation instability is related to the asymmetric response of background leptons and ions to the lepton beam current and can amplify the magnetic field strength and coherence length by orders of magnitude.
We find that this instability operates efficiently over a wide range of conditions and beam-plasma compositions, including pure electron-positron (pair) beams and pair-ion beams with a variable level of pair loading.
In all these cases, this instability can drive a significant energy asymmetry between positively and negatively charged beam particles as it preferentially decelerates electrons.

This instability can have important implications for the magnetization of the precursor of collisionless shocks, including pair-loaded relativistic shocks relevant to GRBs. It will enable a dilute beam of shock accelerated particles to drive near equipartition magnetic fields far ahead of the shock, where plasma microinstabilities are very inefficient. These large-scale magnetic fields are then expected to be advected towards the shock, modifying its structure and affecting nonthermal particle acceleration, radiation emission, and the magnetic field decay in the downstream region. Furthermore, the resulting energy asymmetry could potentially contribute to the overabundance of ions in the accelerated cosmic rays and could also be important for observed matter-antimatter asymmetries, including the galactic positron excess.

Finally, it would be interesting for future work to consider the possibility of studying this cavitation instability in laboratory experiments. Recent work has explored the study of the interplay between oblique and Weibel-type microinstabilities using either electron or electron-positron beams based on conventional RF accelerators \citep{shukla18,arrowsmith21,claveria21}. By considering configurations that would enable significantly denser and/or larger beams to be produced, such as those using picosecond kJ-class laser pulses \citep{shaw18}, it may be possible to probe magnetic field amplification on the longer temporal and spatial scales associated with the cavitation instability.

\acknowledgments
This work was supported by the U.S. Department of Energy SLAC Contract No. DEAC02-76SF00515, by the U.S. DOE FES under FWP 100742, and by the DOE NNSA Laboratory Residency Graduate Fellowship (LRGF) under grant DE-NA0003960. The authors acknowledge the OSIRIS Consortium, consisting of UCLA and IST (Portugal) for the use of the OSIRIS 4.0 framework.
Simulations were performed at Cori (NERSC) through an ALCC computational grant.

\facilities{Cori (NERSC)}

\appendix

\section{Cavitation instability growth rate}\label{appendix:growthrate}

Here we describe in more detail the derivation of the growth rate of the cavitation instability, followed by the analysis of several limiting cases. The mass density which must be expelled from the cavity is the sum of the mass densities of the beam ions, beam positrons, background electrons, background positrons, and all background ions except those with density $\leq n_{be^-}$ left in the cavity to charge-neutralize the beam electrons. This sum produces the mass density
\begin{align}
    \rho_w = m_i\left(\textrm{max}\{n_{0i}-n_{be^-},0\} + \gamma_{bi} n_{bi} \right) + m_e \left(n_{0e^-} + n_{0e^+} + \gamma_{be} n_{be^+} \right).
\end{align}
Substituting for the densities as a function of $Z_\pm$ and using the equipartition between beam kinetic energies $\gamma_{bi} = 1+m_e(\gamma_{be}-1)/m_i$ produces
\begin{align}
    \rho_w = n_{0i} \big\{ m_i\ \textrm{max}\{1-\alpha Z_\pm,\alpha \}  + m_e \left[1+2Z_\pm - \alpha + \alpha \gamma_{be}( Z_\pm+1) \right]\big\}.
\end{align}
By considering the limit $\alpha \ll 1$ we arrive at Eq. \ref{eq:rho} in the main text.

Inserting the full wall mass density $\rho_w$ from Eq. \ref{eq:rho} into the growth rate of Eq. \ref{eq:gr1} produces the full growth rate
\begin{align}
  \frac{\Gamma}{\omega_p} = \alpha \beta_{be}   \left( \frac{
  (1+Z_\pm) m_e\ \textrm{min}\left\{1,1/(Z_\pm\alpha)^2\right\}
  }{
  \delta \big\{ m_i\ \textrm{max}\{1-\alpha Z_\pm,\alpha \}  + m_e \left[1+2Z_\pm - \alpha + \alpha \gamma_{be}( Z_\pm+1) \right]\big\}
  } \right)^{1/2}
\end{align}

Examining this solution in various limits provides valuable insight about the dynamics. For low pair multiplicities $Z_\pm \ll 1/\alpha$, the background ions are still able to charge-neutralize the beam electrons. In this regime, the growth rate becomes
\begin{equation}
  \frac{\Gamma}{\omega_p} \approx \alpha \beta_{be} \sqrt{\frac{m_e (Z_\pm+1)}{\delta \{m_i + m_e[2Z_\pm+\alpha \gamma_{be}(Z_\pm+1)]\}}}.
\end{equation}
For moderate beam energy $1/\alpha \ll \gamma_{be} \ll m_i/m_e$ and moderate pair multiplicity $1 \ll Z_\pm \ll 1/\alpha$ we have
\begin{equation}
  \frac{\Gamma}{\omega_p} \approx \alpha \sqrt{\frac{m_e Z_\pm}{\delta m_i}},
\end{equation}
which shows explicitly the $\Gamma \propto \sqrt{Z_\pm}$ dependence observed in the simulation results of Fig. \hyperref[fig:scalings]{\ref*{fig:scalings}(a)} for $1 < Z_\pm < 10$.

When $Z_\pm=1/\alpha$, the background ions are numerous enough to completely charge-neutralize the beam but do not need to be expelled in the wall, causing the growth rate to reach a maximum. Also considering the limit of ultrarelativistic beam with $\gamma_{be} \gg 2/\alpha$ and $\gamma_{b0} \gg \alpha m_i/m_e$ leads to
\begin{equation}
  \frac{\Gamma}{\omega_p} \approx \sqrt{\frac{\alpha}{\delta \gamma_{be}}},
\end{equation}
which interestingly is, within a factor of order unity, similar to the growth rate of the Weibel instability. This is understood to happen because the cavity dynamics in this regime are determined only by the repulsion of the two beam species.

Finally, when the pair multiplicity increases to $Z_\pm \gg 1/\alpha$, the scarcity of background ions lowers the effective pressure and slows the growth rate as
\begin{equation}
  \frac{\Gamma}{\omega_p} \approx \frac{\beta_{be}}{\sqrt{Z_\pm \delta [m_i\alpha + Z_\pm m_e(2+\alpha \gamma_{be})]}}.
\end{equation}
At moderate beam energies $\gamma_{be} \gg m_i/(Z_\pm m_e)$ and $\gamma_{be} \gg 2/\alpha$, the beam positrons dominate the wall inertia resulting in the growth rate
\begin{equation}
  \frac{\Gamma}{\omega_p} \approx \frac{1}{Z_\pm\sqrt{\delta m_e\alpha \gamma_{be}}}
\end{equation}
which shows the same $\Gamma \propto Z_\pm^{-1}$ scaling observed in Fig. \hyperref[fig:scalings]{\ref*{fig:scalings}(a)} for $Z_\pm > 10$. Remarkably, for very large pair multiplicities, the growth of the cavitation instability actually favors lower beam densities; only a fraction of beam electrons are able to be charge neutralized, yet they must push all of the beam positrons out of the cavity. The growth rate will continue to decrease with $Z_\pm$ until, in the case of a pure pair beam on pair plasma at ($Z_\pm \to \infty$), the cavitation instability will not grow unless an asymmetry between the inertia of electrons and positrons develops due to other processes not considered here.

\section{Periodic Longitudinal Simulations}\label{appendix:longitudinal_simulations}
In simulations where the longitudinal dimension is resolved, the use of fully periodic boundary conditions with a uniform beam can lead to unphysical artifacts often ignored in the literature. In this geometry, the simulation is initialized with overlapping cold beams/plasmas and instabilities start growing throughout the entire simulation domain at the same time. For this reason, causality will artificially limit the longitudinal coherence length of the current filaments produced by the Weibel instability to $L_\parallel \sim c/\Gamma_W$; regions at a larger separation will grow independently from each other. This can have important implications for the growth of the cavitation instability because the electron-driven current filaments need to be longer than $c/\Gamma_\mathrm{C} \gg c/\Gamma_\mathrm{W}$ for the instability to develop (the indices `W' and `C' refer to the Weibel and cavitation instabilities); otherwise, the current from beam positron/ion filaments will disrupt the magnetic field growth. It is thus critical to consider more realistic simulation setups with nonperiodic longitudinal boundary conditions and semi-infinite beams as used in Fig. \ref{fig:semi_infinite}.

\begin{figure}[]
\centering
\includegraphics[width=0.5\linewidth]{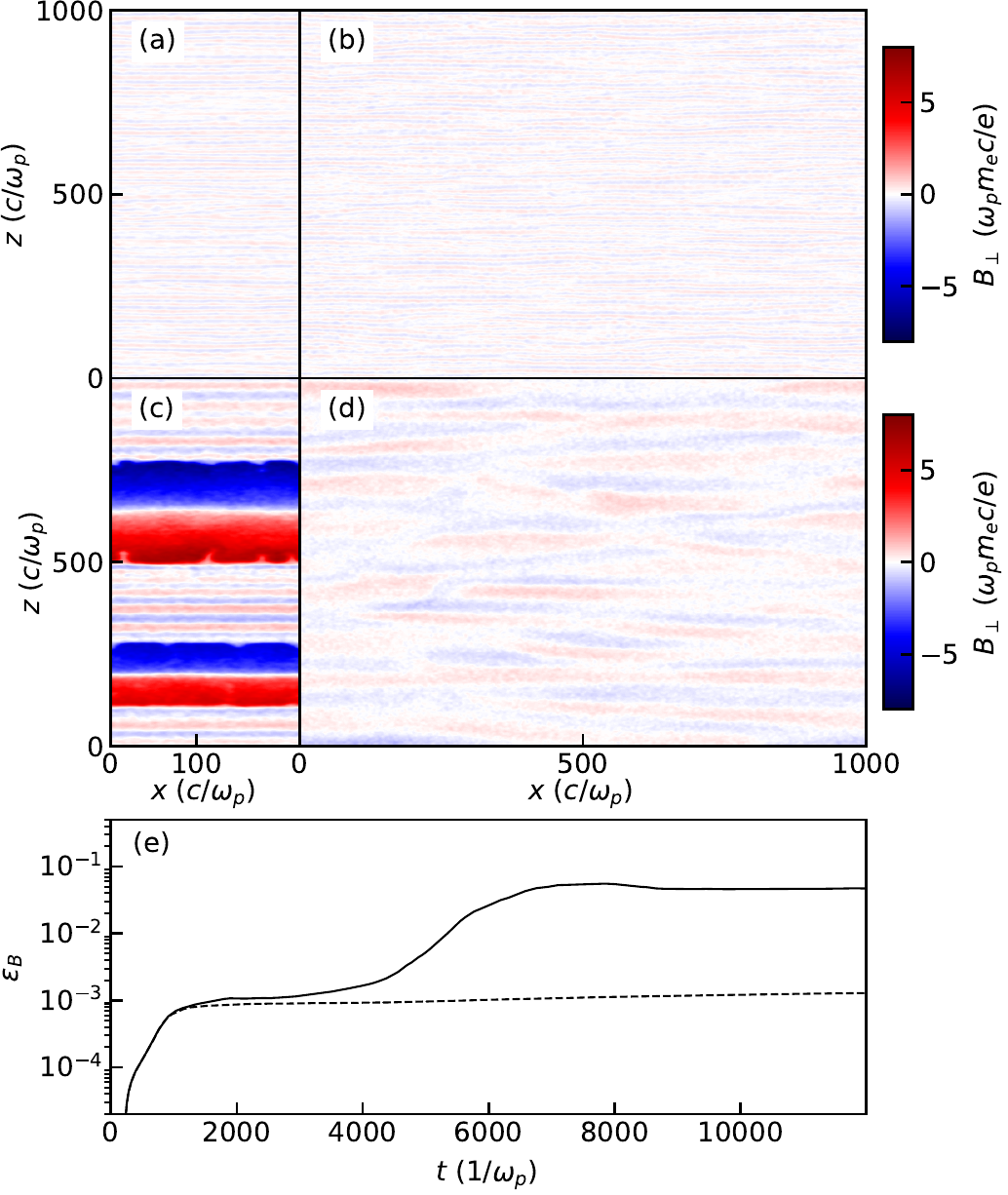}
\caption{\label{fig:infinite_boxes}
Magnetic field amplification from the propagation of a hot pair beam with $\alpha=0.1$, $\langle \gamma_{be} \rangle=1000$, and $T_b^\prime=m_e c^2$ on an electron-ion plasma from 2D fully periodic simulations with longitudinal box size (a,b) $220~c/\omega_p$ and (c,d) $1000~c/\omega_p$. Magnetic field profiles are shown at (a,b) $t=1000~\omega_p^{-1}$, the time of saturation of the Weibel instability, and (c,d) $t=6500~\omega_p^{-1}$, the time of saturation of the cavitation instability. The evolution of the magnetization in the simulations with small (solid) and large (dashed) longitudinal box size is reported in (e).
}
\end{figure}

The effect of periodic boundaries is demonstrated with two simulations of a fully periodic pair beam with $\alpha=0.1$ and $\langle\gamma_{be}\rangle=1000$ propagating in a cold electron-ion plasma. The two simulations have different longitudinal box lengths of $220~c/\omega_p$ and $1000~c/\omega_p$. At early times shown in Fig. \hyperref[fig:infinite_boxes]{\ref*{fig:infinite_boxes}(a,b)} corresponding to saturation of the Weibel instability, the magnetic field profiles are nearly identical. The longitudinal coherence length of the current filaments produced by the Weibel instability in the large simulation is $\sim500~c/\omega_p \sim c/\Gamma_W$. (Note that the observed growth rate of the Weibel instability is slower than the cold limit prediction because the earlier growth of oblique modes heats the beam and background plasma.) This longitudinal coherence length, artificially imposed by the simulation setup, is smaller than the necessary coherence length $c/\Gamma_\mathrm{C} \sim 10^3~c/\omega_p$ required for the cavitation instability. As a result, no growth of the cavitation instability is observed. However, in a simulation using the same parameters but a smaller longitudinal box size, many filaments now extend over the full box size and so have infinite longitudinal coherence length. Indeed, in this case we see the development of the cavitation instability at late times [Fig. \hyperref[fig:infinite_boxes]{\ref*{fig:infinite_boxes}(c,d)}] and in the energy evolution in \hyperref[fig:infinite_boxes]{\ref*{fig:infinite_boxes}(e)} that only the shorter box with infinite-length current filaments is able to trigger the cavitation instability.

\begin{thebibliography}{}
\expandafter\ifx\csname natexlab\endcsname\relax\def\natexlab#1{#1}\fi
\providecommand{\url}[1]{\href{#1}{#1}}
\providecommand{\dodoi}[1]{doi:~\href{http://doi.org/#1}{\nolinkurl{#1}}}
\providecommand{\doeprint}[1]{\href{http://ascl.net/#1}{\nolinkurl{http://ascl.net/#1}}}
\providecommand{\doarXiv}[1]{\href{https://arxiv.org/abs/#1}{\nolinkurl{https://arxiv.org/abs/#1}}}

\bibitem[{Arrowsmith {et~al.}(2021)Arrowsmith, Shukla, Charitonidis, Boni,
  Chen, Davenne, Dyson, Froula, Gudmundsson, Huffman, Kadi, Reville,
  Richardson, Sarkar, Shaw, Silva, Simon, Trines, Bingham, \&
  Gregori}]{arrowsmith21}
Arrowsmith, C.~D., Shukla, N., Charitonidis, N., {et~al.} 2021, Phys. Rev.
  Research, 3, 023103.
\newblock \url{https://doi.org/10.1103/PhysRevResearch.3.023103}

\bibitem[{Beloborodov(2002)}]{beloborodov02}
Beloborodov, A.~M. 2002, Ap. J., 565, 808, \dodoi{10.1086/324195}

\bibitem[{Bresci {et~al.}(2021)Bresci, Gremillet, \& Lemoine}]{bresci21}
Bresci, V., Gremillet, L., \& Lemoine, M. 2021.
\newblock \doarXiv{2111.04651}

\bibitem[{Bret {et~al.}(2010)Bret, Gremillet, \& Dieckmann}]{bret10}
Bret, A., Gremillet, L., \& Dieckmann, M.~E. 2010, Phys. Plasmas, 17, 120501.
\newblock \url{https://doi.org/10.1063/1.3514586}

\bibitem[{Chang {et~al.}(2008)Chang, Spitkovsky, \& Arons}]{chang08}
Chang, P., Spitkovsky, A., \& Arons, J. 2008, Astrophys. J., 674, 378.
\newblock \url{https://doi.org/10.1086/524764}

\bibitem[{Covino {et~al.}(1999)Covino, Lazzati, Ghisellini, Saracco, Campana,
  Chincarini, Serego, Cimatti, Vanzi, Pasquini, Haardt, Israel, Stella, , \&
  Vietri}]{covino99}
Covino, S., Lazzati, D., Ghisellini, G., {et~al.} 1999, Astronom. Astrophys.,
  348, L1

\bibitem[{Davidson {et~al.}(1972)Davidson, Hammer, Haber, \&
  Wagner}]{davidson72}
Davidson, R.~C., Hammer, D.~A., Haber, I., \& Wagner, C.~E. 1972, Phys. Fluids,
  15, 317.
\newblock \url{https://doi.org/10.1063/1.1693910}

\bibitem[{Fiuza {et~al.}(2012)Fiuza, Fonseca, Tonge, Mori, \& Silva}]{fiuza12}
Fiuza, F., Fonseca, R.~A., Tonge, J., Mori, W.~B., \& Silva, L.~O. 2012, Phys.
  Rev. Lett., 108, 235004.
\newblock \url{http://dx.doi.org/10.1103/PhysRevLett.108.235004}

\bibitem[{Fonseca {et~al.}(2008)Fonseca, Martins, Silva, Tonge, Tsung, \&
  Mori}]{fonseca08}
Fonseca, R.~A., Martins, S.~F., Silva, L.~O., {et~al.} 2008, Plasma Phys.
  Control. Fusion, 50, 124034.
\newblock \url{https://doi.org/10.1088/0741-3335/50/12/124034}

\bibitem[{Fonseca {et~al.}(2002)Fonseca, Silva, Tsung, Decyk, Lu, Ren, Mori,
  Deng, Lee, Katsouleas, \& Adam}]{fonseca02}
Fonseca, R.~A., Silva, L.~O., Tsung, F.~S., {et~al.} 2002, Lect. Notes Comput.
  Sc., 2331, 342.
\newblock \url{https://doi.org/10.1007/3-540-47789-6_36}

\bibitem[{Fried(1959)}]{fried59}
Fried, B.~D. 1959, TRW Space Technology Labs Los Angeles, Calif.

\bibitem[{Gill \& Granot(2020)}]{gill20}
Gill, R., \& Granot, J. 2020, Mon. Not. R. Astron. Soc., 491, 5815.
\newblock \url{https://doi.org/10.1093/mnras/stz3340}

\bibitem[{Gruzinov \& Waxman(1999)}]{gruzinov99}
Gruzinov, A., \& Waxman, E. 1999, Astrophys. J., 511, 852.
\newblock \url{https://doi.org/10.1086/306720}

\bibitem[{Honda {et~al.}(2000)Honda, Meyer{-}ter{-}Vehn, \&
  Pukhov}]{honda00pop}
Honda, M., Meyer{-}ter{-}Vehn, J., \& Pukhov, A. 2000, Phys. Plasmas, 7, 1302,
  \dodoi{10.1063/1.873941}

\bibitem[{Keshet {et~al.}(2009)Keshet, Katz, Spitkovsky, \& Waxman}]{keshet09}
Keshet, U., Katz, B., Spitkovsky, A., \& Waxman, E. 2009, Astrophys. J., 693,
  L127.
\newblock \url{https://doi.org/10.1088/0004-637X/693/2/L127}

\bibitem[{Lemoine(2015)}]{lemoine15}
Lemoine, M. 2015, J. Plasma Phys., 81, 455810101.
\newblock \url{https://doi.org/10.1017/S0022377814000920}

\bibitem[{Lemoine {et~al.}(2019)Lemoine, Gremillet, Pelletier, \&
  Vanthieghem}]{lemoine19PRL}
Lemoine, M., Gremillet, L., Pelletier, G., \& Vanthieghem, A. 2019, Phys. Rev.
  Lett., 123, 035101.
\newblock \url{https://doi.org/10.1103/PhysRevLett.123.035101}

\bibitem[{Martins {et~al.}(2009)Martins, Fonseca, Silva, \& Mori}]{martins09}
Martins, S.~F., Fonseca, R.~A., Silva, L.~O., \& Mori, W.~B. 2009, Astrophys.
  J. Lett., 695, L189.
\newblock \url{http://dx.doi.org/10.1088/0004-637X/695/2/L189}

\bibitem[{Medvedev \& Loeb(1999)}]{medvedev99}
Medvedev, M.~V., \& Loeb, A. 1999, Astrophys. J., 526, 697

\bibitem[{M{\'e}sz{\'a}ros {et~al.}(2001)M{\'e}sz{\'a}ros, Ramirez-Ruiz, \&
  Rees}]{meszaros01}
M{\'e}sz{\'a}ros, P., Ramirez-Ruiz, E., \& Rees, M.~J. 2001, Astrophys. J.,
  554, 660.
\newblock \url{https://doi.org/10.1086/321404}

\bibitem[{Naseri {et~al.}(2018)Naseri, Bochkarev, Ruan, Bychenkov, Khudik, \&
  Shvets}]{naseri18}
Naseri, N., Bochkarev, S.~G., Ruan, P., {et~al.} 2018, Phys. Plasmas, 25,
  012118, \dodoi{10.1063/1.5008278}

\bibitem[{Peterson {et~al.}(2021)Peterson, Glenzer, \& Fiuza}]{peterson21}
Peterson, J.~R., Glenzer, S., \& Fiuza, F. 2021, Phys. Rev. Lett., 126, 215101.
\newblock \url{https://doi.org/10.1103/PhysRevLett.126.215101}

\bibitem[{Ramirez-Ruiz {et~al.}(2007)Ramirez-Ruiz, Nishikawa, \&
  Hededal}]{ramirez-ruiz07}
Ramirez-Ruiz, E., Nishikawa, K.-I., \& Hededal, C.~B. 2007, Ap. J., 671, 1877,
  \dodoi{10.1086/522072}

\bibitem[{Ruyer {et~al.}(2015)Ruyer, Gremillet, \& Bonnaud}]{ruyer15}
Ruyer, C., Gremillet, L., \& Bonnaud, G. 2015, Phys. Plasmas, 22, 082107,
  \dodoi{10.1063/1.4928096}

\bibitem[{{San Miguel Claveria} {et~al.}(2021){San Miguel Claveria}, Davoine,
  Peterson, Gilljohann, Andriyash, Ariniello, Ekerfelt, Emma, Faure, Gessner,
  Hogan, Joshi, Keitel, Knetsch, Kononenko, Litos, Mankovska, Marsh, Matheron,
  Nie, O'Shea, Storey, Vafaei-Najafabadi, Wu, Xu, Yan, Zhang, Tamburini, Fiuza,
  Gremillet, \& Corde}]{claveria21}
{San Miguel Claveria}, P., Davoine, X., Peterson, J.~R., {et~al.} 2021.
\newblock \doarXiv{2106.11625}

\bibitem[{Shaw {et~al.}(2021)Shaw, Romo-Gonzalez, Lemos, King, Bruhaug, Miller,
  Dorrer, Kruschwitz, Waxer, Williams, Ambat, McKie, Sinclair, Mori, Joshi,
  Chen, Palastro, Albert, \& Froula}]{shaw18}
Shaw, J.~L., Romo-Gonzalez, M.~A., Lemos, N., {et~al.} 2021, Sci. Rep., 11,
  7498.
\newblock \url{https://doi.org/10.1038/s41598-021-86523-5}

\bibitem[{Shukla {et~al.}(2018)Shukla, Viera, Muggli, Sarri, Fonseca, \&
  Silva}]{shukla18}
Shukla, N., Viera, J., Muggli, P., {et~al.} 2018, J. Plasma Phys., 84,
  905840302.
\newblock \url{https://doi.org/10.1017/S0022377818000417}

\bibitem[{Silva {et~al.}(2003)Silva, Fonseca, Tonge, Dawson, Mori, \&
  Medvedev}]{silva03}
Silva, L.~O., Fonseca, R.~A., Tonge, J.~W., {et~al.} 2003, Ap. J., 596, L121,
  \dodoi{10.1086/379156}

\bibitem[{Silva {et~al.}(2002)Silva, Fonseca, Tonge, Mori, \& Dawson}]{silva02}
Silva, L.~O., Fonseca, R.~A., Tonge, J.~W., Mori, W.~B., \& Dawson, J.~M. 2002,
  Phys. Plasmas, 9, 2458.
\newblock \url{https://doi.org/10.1063/1.1476004}

\bibitem[{Sironi \& Giannios(2014)}]{sironi14}
Sironi, L., \& Giannios, D. 2014, Ap. J., 787, 49.
\newblock \url{http://dx.doi.org/10.1088/0004-637X/787/1/49}

\bibitem[{Sironi {et~al.}(2013)Sironi, Spitkovsky, \& Arons}]{sironi13}
Sironi, L., Spitkovsky, A., \& Arons, J. 2013, Astrophys. J., 771, 54.
\newblock \url{https://doi.org/10.1088/0004-637X/771/1/54}

\bibitem[{Spitkovsky(2007)}]{spitkovsky07}
Spitkovsky, A. 2007, Astrophys. J., 673, L39.
\newblock \url{https://doi.org/10.1086/527374}

\bibitem[{Spitkovsky(2008)}]{spitkovsky08}
---. 2008, Astrophys. J., 682, L5.
\newblock \url{https://doi.org/10.1086/590248}

\bibitem[{Steele {et~al.}(2009)Steele, Mundell, Smith, Kobayashi, \&
  Guidorzi}]{steele09}
Steele, I.~A., Mundell, C.~G., Smith, R.~J., Kobayashi, S., \& Guidorzi, C.
  2009, Nature, 462, 767, \dodoi{10.1038/nature08590}

\bibitem[{Thompson \& Madau(2000)}]{thompson00}
Thompson, C., \& Madau, P. 2000, Astrophys. J., 538, 105.
\newblock \url{https://doi.org/10.1086/309100}

\bibitem[{Weibel(1959)}]{weibel59}
Weibel, E. 1959, Phys. Rev. Lett., 2, 83.
\newblock \url{https://doi.org/10.1103/PhysRevLett.2.83}

\end{thebibliography}
\end{document}